\documentclass[%
 reprint,
 amsmath,amssymb,
 aps,
 prl
]{revtex4-2}

\usepackage{graphicx}
\usepackage{dcolumn}
\usepackage{bm}
\usepackage{hyperref}

\begin{document}

\preprint{APS/123-QED}

\title{Electron heating at quasi-perpendicular collisionless shocks}
\newcommand{\jgr}{Journal of Geophysical Research}

\author{Ahmad Lalti}
\email{ahmad.lalti@irfu.se}
 \affiliation{%
Swedish Institute of Space Physics, Uppsala, Sweden}
\affiliation{Department of Physics and Astronomy, Space and Plasma Physics, Uppsala University, Sweden 
}
\author{Yuri V. Khotyaintsev}
\affiliation{
Swedish Institute of Space Physics, Uppsala, Sweden
}%
\author{Daniel B. Graham}
\affiliation{
Swedish Institute of Space Physics, Uppsala, Sweden
}%

\date{\today}
\begin{abstract}
Adiabatic and non-adiabatic electron dynamics have been proposed to explain electron heating across collisionless shocks. We analyze the evolution of the suprathermal electrons across 310 quasi-perpendicular shocks with $1.7<M_A<48$ using in-situ measurements. We show that the electron heating mechanism shifts from predominantly adiabatic to non-adiabatic for the Alfv\'enic Mach number in the de Hoffman-Teller $\gtrsim 30$ with the latter constituting 48\% of the analyzed shocks. The observed non-adiabatic heating is consistent with the stochastic shock drift acceleration mechanism.

\end{abstract}

\maketitle

\emph{Introduction}.--
Collisionless shocks are ubiquitous in space, yet their dynamics remain an active field of research. One of the remaining open questions is that of electron heating and energization across the shock. Two competing frameworks have been suggested since the early days of collisionless shocks physics, both of which have theoretical, observational and simulation studies published to support them. 

The first framework revolves around the adiabatic electron dynamics across the shock where the electrons remain magnetized and the drift approximation can be used \cite{goodrich1984adiabatic,schwartz1988electron,savoini1994electron,scudder1995review}. As the electrons cross the shock, they are accelerated into the downstream by the cross-shock electric field and gain the energy of the cross-shock potential in the de Hoffmann-Teller (HT) reference frame ($\Delta \Phi_{HT} = \Phi_{down} - \Phi_{up}$) \cite{thomsen1987strong,schwartz1988electron,scudder1986highbeta_b,scudder1986highbeta_c,hull2000electron}. In addition, to conserve their first adiabatic invariant, the perpendicular energy of the electrons increases in response to the increase in the magnetic field magnitude $|\vec{B}|$. Such adiabatic process creates large gradients in the electron distribution, rendering it unstable to various micro-instabilities, which subsequently irreversibly smoothen those gradients and form a flat-topped distribution \cite{feldman1983electron,scudder1986highbeta_c,hull1998electron,hull2001electron,lefebvre2007electron}. In other words, electron heating in this framework is controlled by adiabatic-reversible interactions with the DC-fields of the shock, with secondary non-adiabatic irreversible effects coming from microinstabilities. 

The second framework revolves around the non-adiabatic electron dynamics across the shock, where wave-particle interaction between electrons and various plasma wave modes excited in the shock ramp non-adiabatically heats up and energize the electrons across the shock \cite{sagdeev1966cooperative,wu1984microinstabilities,papadopoulos1985microinstabilities,wilson2014quantified1,wilson2014quantified2}. Wave-particle interaction can break the adiabaticity of electrons in various ways. To start with, electromagnetic and electrostatic waves have been shown both using theory \cite{cargill1988mechanism,hoshino2002nonthermal,katou2019theory,amano2022theory,kamaletdinov2022quantifying}and observations \cite{amano2020observational} to be able to scatter the electrons and trap them within the shock ramp itself. Trapped electrons can subsequently become energized by the convective electric field shaping the downstream suprathermal tail of the distribution and heating the electrons. In addition, it has been suggested that structures with large electric field gradients can demagnetize and energize the electrons \cite{cole1976effects,balikhin1998study,see2013non}. Such large gradient structures have been observed using in-situ measurement across Earth's bow shock \cite{bale2002electrostatic,schwartz2011electron,wang2020electrostatic,vasko2022ion} and it was shown that they indeed can affect the electrons' dynamics \cite{chen2018electron,vasko2018solitary,kamaletdinov2022quantifying}.

In this letter, we analyze 310 quasi-perpendicular bow shock crossings observed by the Magnetospheric MultiScale (MMS) spacecraft \cite{MMS_overview} to understand which of the two frameworks is more suitable to describe the electrons' heating and energization across collisionless shocks.

\begin{figure*}
    \centering
    \includegraphics[width=17.6cm]{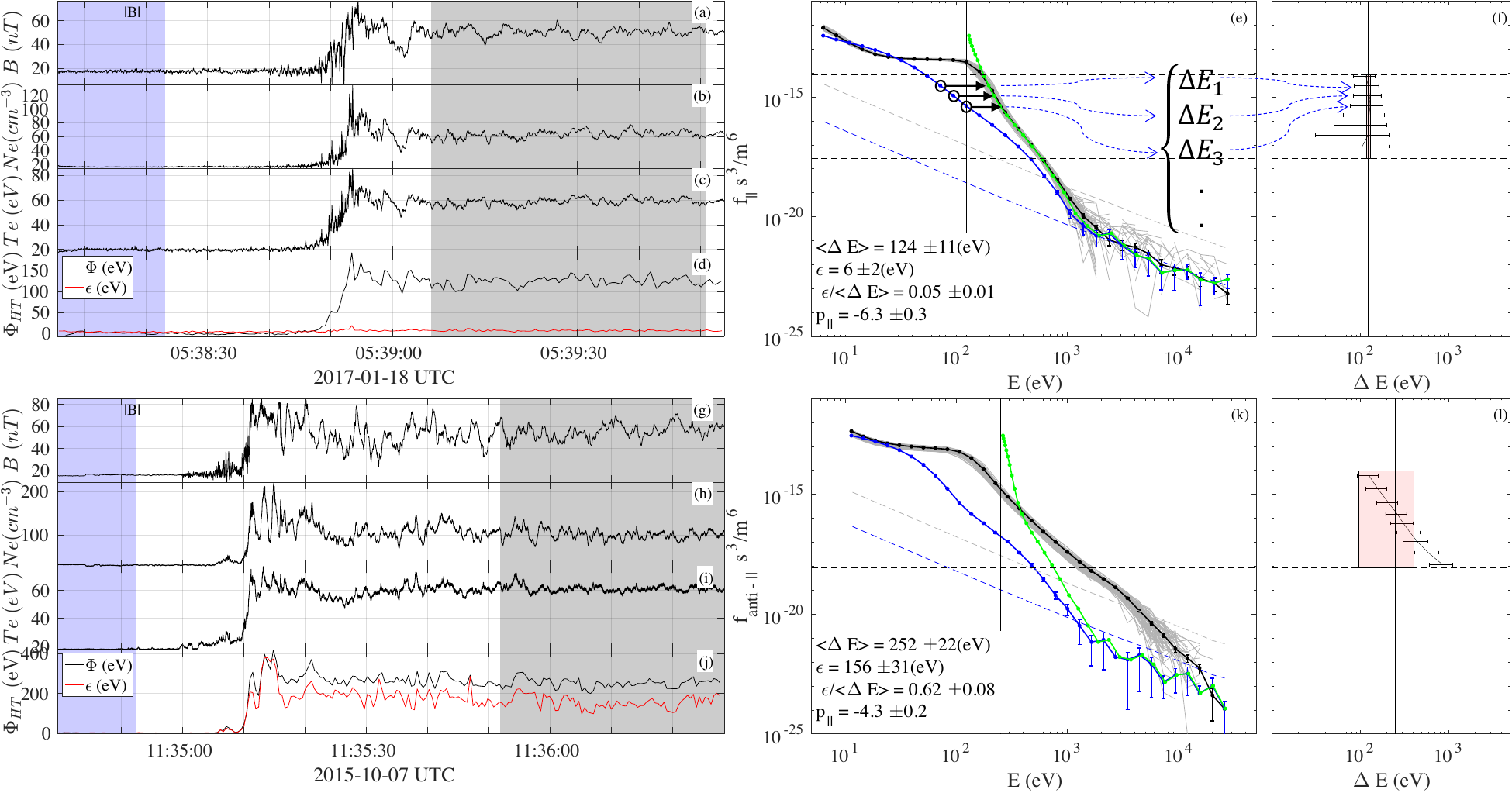}
    \caption{Two example shock crossings. Example 1 is shown in panels a-f, while example 2 is in panels g-l. Panels (a,g), (b,h), and (c,i) show, respectively, the magnetic field magnitude, electron density, and electron temperature. Panels (d,j) show in black $\Delta \Phi$ and in red $\epsilon$. Panels (e,k) show the upstream (blue), downstream (gray and black), and Liouville mapped (green) parallel cuts in the distribution function $f_{\parallel}$. The blue and gray dashed lines are the 1-count levels for the blue and gray distributions. The black dashed lines are the region on which the analysis is applied, and the vertical black line is at $E = \Delta \Phi$. Panels (f,l) shows $f_{\parallel}$ versus $\Delta E$ with the pink shaded box representing interval $\Delta \Phi \pm \epsilon$.}
    \label{fig:figad}
\end{figure*}

\emph{Method}.-- 
We use the MMS bow shock crossings database \cite{lalti2022database} to select quasi-perpendicular shocks ($45^\circ < \theta_{Bn} < 90^\circ$), with relatively stable upstream conditions, i.e. small variations in the solar wind magnetic field and density. We focus on shocks for which MMS operated in burst mode (high sampling frequency data acquisition mode). We find 310 shocks satisfying those conditions. For each one of those shocks we select the upstream and downstream intervals, and using magnetic field data from the FluxGate Magnetometer (FGM) \cite{russell2016magnetospheric}, and electron and ion moments data provided by the Dual-Electron and Dual-Ion spectrometers (DES/DIS) of the Fast Plasma Investigation (FPI) instrument suite \cite{pollock2016fast} we obtain the main shock parameters ($M_{A-NIF}$ and  $M_{f-NIF}$, the Alfv\'enic and fast mode Mach numbers in the normal incidance frame (NIF), and electron and ion betas $\beta_{e/i}$). The local shock normal was determined using the mixed-mode 3 method based on the coplanarity theorem \cite{schwartz1998}.

An example MMS shock crossing with $M_{A-NIF} = 4.2$ and $\theta_{Bn} = 69^\circ$ (shock 1) is shown in Fig. \ref{fig:figad}. The magnetic field magnitude, the electron density, and electron temperature show a clear quasi-perpendicular shock signature, i.e., upstream, foot, ramp, and downstream \cite{balogh2013physics}, see Fig.~\ref{fig:figad}a-c, where the upstream and downstream intervals selected have been highlighted in blue and black respectively.

We use the electron distribution functions measured by the dual electron spectrometers (DES) from FPI at 30 ms cadence. Those are all-sky distributions binned on a spherical grid covering 4$\pi$ sr field of view and energies from 10 eV to 30 keV at a $\delta E/E$ of $\sim 14\%$ \cite{pollock2016fast}. We analyze the distribution in the magnetic field aligned coordinate system $f\left(E_{\parallel},E_{\perp}\right)$, with $E_{\parallel}$ and $E_{\perp}$ being the parallel and perpendicular energies to the background magnetic field. We show parallel cuts of the distributions, $f_{\parallel} = f\left(E_{\parallel},E_{\perp} = 0\right)$, for the example shock 1 in Fig.~\ref{fig:figad}e. The upstream distribution (blue curve) is obtained by averaging the 3D distribution over the upstream region (blue-shaded region) and then taking the parallel cut. The error bars represent the uncertainty from the counting statistics ($\sigma_f$). The 1-count level $f_{1c}$ for that averaging interval, i.e., the distribution value obtained if the sensor had registered a single electron count, is shown by blue dashed line. Note that the larger the number of distributions we average over, the smaller the 1-count level is. The gray curves represent all $f_{\parallel}$ within the downstream interval (black shaded area); the black curve is their average. The individual $f_{\parallel}$ are produced by averaging over 16 observed 30-ms distributions (the gray dashed line shows the corresponding 1-count level). The downstream distribution has a clear flat-topped shape, consisting of a flat part at energies $\lesssim 10^{2}$ eV and a power-law part at energies  $\gtrsim 10^{2}$ eV. At even higher energies, the distribution becomes noisy due to low counting statistics, $f_{\parallel} \lesssim f_{1c}$.

We test if the behavior of the supra-thermal electrons (the power-law part of the distribution) across the shock is adiabatic using a modified approach to that in Refs.~\cite{hull1998electron,lefebvre2007electron,schwartz2021evaluating,johlander2023electron}. We assume that the drift approximation is valid and that electrons conserve their first adiabatic invariant and total energy. Under these assumptions, Liouville's theorem holds, which states that the phase space density of electrons $f\left(E_{\parallel},E_{\perp}\right)$ is conserved along characteristics of the motion. For the sake of our analysis, it is sufficient to focus on parallel cuts in the distribution $f_{\parallel}$. Liouville's theorem states that $f\left(E_{\parallel-up}\right) = f\left(E_{\parallel-down}\right)$, and by using conservation of energy we can write
\begin{equation}
    E_{\parallel-down} = E_{\parallel-up}+e\Delta\Phi_{HT},
    \label{eq:delta_phi}
\end{equation}
with $e$ being the electron charge.
Subsequently, we find $\Delta\Phi_{HT}$ that best maps the observed reference upstream distribution to the downstream.

In theory, the analysis above should be done in the HT frame. However, the measured distributions are binned on a spherical grid that is logarithmically spaced in energy, and transforming them to the HT frame can introduce very large uncertainties into the distributions, especially for highly perpendicular shocks. Therefore, we perform the analysis in the original spacecraft (SC) frame, and as we focus on electrons with suprathermal energies, the analysis can be done in the SC frame without introducing significant uncertainty to the results (see Supplemental Material for details \cite{supplemental_material}).

Within the adiabatic framework, the flat part of the flat-topped distribution is formed by non-adiabatic wave-particle interactions \cite{scudder1995review}. In contrast, for the suprathermal electrons with energies beyond the knee of the downstream flat-topped distribution (the power-law part of the distribution), adiabatic dynamics dominate, and Liouville theorem is valid. We select the upper bound of the distribution $f_{max}$  to be beyond the knee of the downstream distribution. Furthermore, we make sure that the lowest value in the measured distribution included in the analysis $f_{min}$ satisfies the condition $f_{1c}<f_{min}-3\sigma_{f_{min}}$. The black dashed horizontal lines in panel (e) of Fig. \ref{fig:figad} indicate the $\left[f_{min},\; f_{max}\right]$ interval for this specific shock over which we apply our analysis. Within that interval and for each value of the reference distribution $f_{\parallel-up,i}$ corresponding to the different energy bins ($i$ is the index of each energy bin), we calculate the energy $\Delta E_{i}$ necessary to map $f_{\parallel-up,i}$ to the downstream distribution; this is represented schematically in panels (e-f). In panel (f) we plot $f_{\parallel-up}$ versus $\Delta E$ with the error bars representing the uncertainty in $\Delta E$ coming from the counting statistics and the finite energy bin width ($\delta E/E \sim 14\%$).

To estimate the cross-shock potential we calculate the weighted mean of the set of $\Delta E_i$ with the weights $w_i$ calculated from the uncertainties, $\Delta \Phi_{HT} = \left<w_i \Delta E_i\right>$. We use that value of $\Delta \Phi_{HT}$ to map the reference distribution shown in blue in panel (e). Comparing the mapped (green) to the observed (black) distributions within the interval $\left[f_{min},\; f_{max}\right]$, we find good agreement between them, indicating that the adiabatic framework of electron heating describes the electron behavior for this shock. 

To quantify the adiabaticity of the electrons, we calculate the weighted standard deviation of $\Delta E_i$, $\epsilon = std\left(w_i \Delta E_i\right)$ and we normalize it to $\Delta \Phi_{HT}$. Small $\epsilon/\Delta \Phi_{HT}$ indicates a good overlap between the mapped and observed distributions and hence adiabatic behavior. As mentioned above, the FPI/DES instrument has $\delta E/E \sim 0.14$, which results in $\epsilon/\Delta \Phi_{HT}$ being different from zero even for shocks with the pure adiabatic behavior. Therefore, the values of $\frac{\epsilon}{\Delta \Phi_{HT}}<2\cdot\frac{\delta E}{E} \sim 0.28$ are within the instrument uncertainty.

We map the reference upstream distribution to all the other distributions and calculate $\Delta \Phi_{HT}$ and $\epsilon$. Fig. \ref{fig:figad}d shows $\Delta \Phi_{HT}$ in black and $\epsilon$ in red as a time series for shock 1. We see that $\Delta \Phi_{HT}$ tracks the electron temperature curve, and $\epsilon$ remains small throughout the event, including downstream of the shock. To capture the time variation in $\Delta \Phi_{HT}$ and $\epsilon$, we calculate their mean and standard deviation within the downstream (black-shaded) interval shown in the lower left corner of the panel (e). We can see that $\frac{\epsilon}{\Delta \Phi} = 0.05 \pm 0.01< 0.28$, i.e. within the range attributed to instrument uncertainty, which provides quantitative confirmation of the adiabatic electron behavior for this shock. 

We present a second example of a shock crossing with $M_{A-NIF} = 6.7$ and $\theta_{Bn} = 86.5^\circ$  (shock 2) Fig. \ref{fig:figad}g-l. The magnetic field, electron density, and temperature time series all show the same typical signature of a quasi-perpendicular shock as in shock 1. From panels (k-l), we see that the Liouville mapped distribution (green) deviates significantly from the observed distribution (black) within the interval $\left[f_{min},\; f_{max}\right]$. This is reflected in the large slope in the $f_{\parallel-up}$ versus $\Delta E$ curve shown in panel (l) and in the large value of $\frac{\epsilon}{\Delta \Phi} = 0.62\pm 0.08$. This indicates that non-adiabatic processes play a major role in the dynamics of suprathermal electrons for this shock. Another implication of the large values of $\frac{\epsilon}{\Delta \Phi}$ is that the Liouville mapping estimate of $\Delta \Phi_{HT}$ becomes unreliable.

Another way to differentiate between the heating mechanisms would be to investigate the exact shape of the downstream distribution. At suprathermal energies, the distribution can be approximated by a power-law: $f_{\parallel}\sim E^{p_{\parallel}}$, with $p_{\parallel}$ being the spectral index. 
We fit the observed downstream $f_{\parallel}$ in the interval $f \in \left[f_{min}, \; f_{max}\right]$ for the example shocks 1 and 2. Similar to $\epsilon$ and $\Delta \Phi$, we calculate the mean and standard deviation in $p_{\parallel}$ for the downstream interval. We find $p_{\parallel} = -6.3 \pm 0.3$ for shock 1 (Fig.~\ref{fig:figad}e) and $p_{\parallel} = -4.3 \pm 0.2$ for shock 2 (Fig.~\ref{fig:figad}k). So, the shock with adiabatic behavior (shock 1) has a softer spectrum.

\begin{figure}[h]
    \centering
    \includegraphics[width=8.6cm]{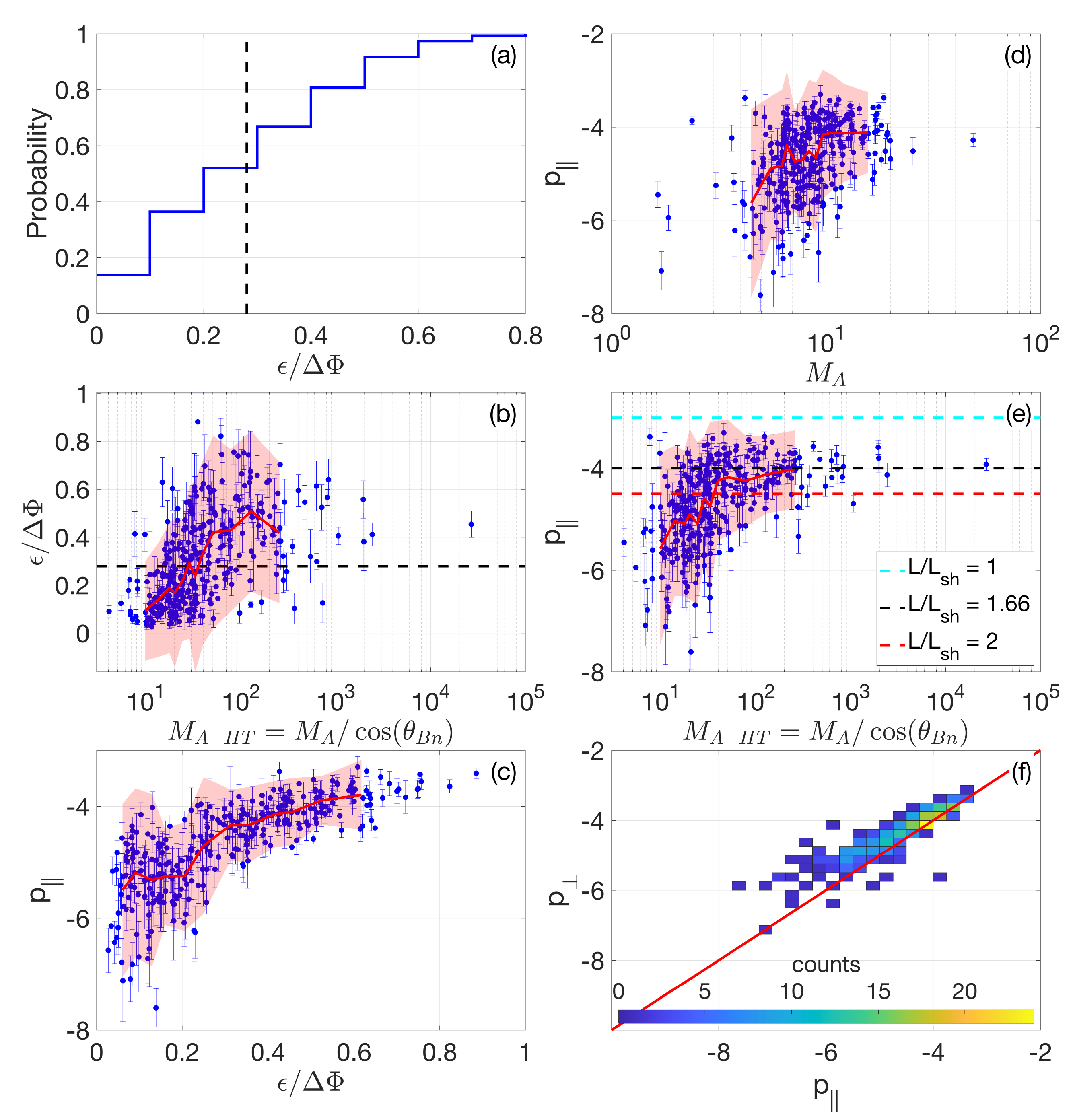}
    \caption{Panel (a) shows the cumulative distribution function of $\epsilon/\Delta \Phi$, panel (b) shows $\epsilon/\Delta \Phi$ versus $M_{A-HT}$, panel (c) shows $\epsilon/\Delta \Phi$ versus $p_{\parallel}$, panel (d-e) shows $p_{\parallel}$ versus $M_A$ and $M_{A-HT}$ respectively, and panel (f) shows the 2d histogram of $p_{\perp}$ versus $p_{\parallel}$ with the colorbar representing the number of shocks in each bin. The red line and the pink shaded area in panels (b-e) represent the median along with the 95\% confidence interval for the plotted data in each panel, binned in bins containing 25 data points each.}
    \label{fig:figstat}
\end{figure}

\emph{Results}.-- The results of the above analysis applied to 310 shocks in our list, are presented in Fig. \ref{fig:figstat}. We find that $50\%$ of the shocks have $\frac{\epsilon}{\Delta \Phi}>0.28$ (Fig. \ref{fig:figstat}a), and hence for those shocks, non-adiabatic processes play a measurable role in the dynamics of suprathermal electrons. This indicates that despite Earth's bow shock being characterized with relatively low $M_{A-NIF}\sim 5-15$ \cite{lalti2022database}, the electron dynamics can deviate from being adiabatic. 

We further explore the dependence of $\frac{\epsilon}{\Delta \Phi}$ on the various shock parameter and find the strongest correlation with $M_{A-HT} = M_{A-NIF}/\cos(\theta_{Bn})$ (Fig. \ref{fig:figstat}b). For $M_{A-HT} \lesssim 20$  we find small values of the non-adiabaticity measure, $\frac{\epsilon}{\Delta \Phi}\lesssim 0.1$ which indicates that non-adiabatic effects are relatively unimportant. The scatter in $\frac{\epsilon}{\Delta \Phi}$ increases with $M_{A-HT}$, so individual shocks with high $M_{A-HT}$ can still have small $\frac{\epsilon}{\Delta \Phi} \gtrsim 0.1$, but statistically $\frac{\epsilon}{\Delta \Phi}$ increases with $M_{A-HT}$. For $M_{A-HT} \gtrsim 30$, we find the median value of $\frac{\epsilon}{\Delta \Phi}>0.28$ indicating a transition to shocks where non-adiabatic effects become measurable.

We also find a clear positive correlation between $p_{\parallel}$ and $\frac{\epsilon}{\Delta \Phi}$, see Fig.\ref{fig:figstat}c. 
This indicates that harder spectra $p_{\parallel} \gtrsim -4.5$ tend to be produced by shocks with $\frac{\epsilon}{\Delta \Phi}\gtrsim0.3$. We further explore the dependence of $p_{\parallel}$ on various shock parameters, and we find a positive correlation with both $M_{A-NIF}$ (Fig. \ref{fig:figstat}d) and $\theta_{Bn}$ (not shown). Similar correlations for the spectral index but for the omni-direction distribution were reported in Ref.~\onlinecite{oka2006whistler}. We find that the strongest correlation of $p_{\parallel}$ is with $M_{A-HT}$ (Fig. \ref{fig:figstat}e) which combines the dependencies on $M_{A-NIF}$ and $\theta_{Bn}$. At smaller values of $M_{A-HT} \lesssim 30$, the spectrum is softer ($p_{\parallel} \lesssim -4.5$), and as $M_{A-HT}$ increases, the spectrum becomes harder, and the spectral index approaches a plateau at $p_{\parallel}\sim -4$.

We compare the observed behavior of the spectral index to the prediction of the Stochastic Shock Drift Acceleration (SSDA) theory \cite{katou2019theory,amano2022theory}. In SSDA electrons become trapped within the shock layer by pitch-angle scattering on whistler waves, while trapped, they become energized by the convective electric field. This mechanism is stochastic in nature. Assuming that the trapping efficiency is independent of the energy of the electrons, it yields a power-law distribution for the trapped electrons \cite{oka2018electron,amano2022theory}. The predicted spectral index is given by
\begin{equation}
    p = -\frac{3}{2}\left(1+\frac{L}{L_{sh}}\right),
    \label{eq:ssda}
\end{equation}
where $L$ and $L_{sh}$ is the magnetic gradient and the shock transition region length scales along the magnetic field. This expression depends only on the ratio $\frac{L}{L_{sh}}$, which is of the order unity \cite{katou2019theory}. In Fig.~\ref{fig:figstat}e, we overlay $p$ from Eq.~\ref{eq:ssda} for three different values of $\frac{L}{L_{sh}}$. We find that the plateau in the observed spectral index $p_{\parallel}$ at large $M_{A-HT}$ has a value consistent with the SSDA prediction for $\frac{L}{L_{sh}} = 1.66$. 

A stochastic process should result in a relatively isotropic distribution \cite{amano2020observational}. Taking cuts in the perpendicular direction to the background magnetic field we calculate the spectral index $p_{\perp}$ similar to $p_{\parallel}$. Figure \ref{fig:figstat}f shows a 2D histogram of $p_{\perp}$ versus $p_{\parallel}$, with the red line representing $p_{\perp} = p_{\parallel}$. We see that for softer spectra $p_{\parallel}\lesssim -4.5$, there is slight anisotropy in the distribution with $p_{\perp}>p_{\parallel}$. On the other hand, for harder spectra $p_{\parallel}\gtrsim -4.5$ we find $p_{\perp} \approx p_{\parallel}$, suggesting more isotropic distributions. This further indicates that stochastic dynamics of suprathermal electrons dominate for shocks with harder spectra.

\emph{Discussion.}-- We observe a positive correlation between our non-adiabaticity measure $\epsilon/\Delta \Phi$ and $M_{A-HT}$ (Fig. \ref{fig:figstat}b). By definition, the closer $\epsilon/\Delta \Phi$ is to zero the more dominant the adiabatic electron dynamics is. As $M_{A-HT}$ increases, the contribution of non-adiabatic effects increases (i.e. $\epsilon/\Delta \Phi$ deviates from zero). Shocks with harder spectra $p_{\parallel} \gtrsim -4.5$ approach the SSDA predicted plateau and have $p_{\perp} \approx p_{\parallel}$. This indicates that for those shocks, stochastic electron dynamics, and hence the non-adiabatic electron heating framework, dominates. Such shocks have a non-adiabaticity measure $\epsilon/\Delta \Phi \gtrsim 0.3$. 48\% of the analyzed shocks belong to that category. It is worth noting that shocks with harder spectra are more likely to inject electrons into the Diffusive Shock Acceleration (DSA) process \cite{amano2022nonthermal}.

This result is consistent with previous observational case studies using Liouville mapping to analyze electron heating. By visual inspection of the distributions, it was found that for small $M_{A-NIF}$ and $\theta_{Bn}$ the mapped and observed distributions overlapped very well \cite{hull2001electron,lefebvre2007electron,johlander2023electron}, while a shock with larger $M_{A-NIF}$ and $\theta_{Bn}$ showed significant deviation between the mapped and observed distributions \cite{johlander2023electron}.

Furthermore, from Figs.~\ref{fig:figstat}b and \ref{fig:figstat}e, one can see that the transition between adiabatic and stochastic heating is continuous, and at intermediate values of $M_{A-HT} \sim 30$ and $\epsilon/\Delta \Phi \lesssim 0.3$ both adiabatic and stochastic effects are important. Yet the limits of this intermediate range is not easily discernible from our results due to the large scatter in $\epsilon/\Delta \Phi$. The $\delta E/E \sim 0.14$ of the instrument can contribute to the observed scatter in $\epsilon/\Delta \Phi$. An instrument with smaller $\delta E/E$ is needed to probe the electron dynamics in the regime where both adiabatic and non-adiabatic effects are important. Additionally, this scatter can also be attributed either to large uncertainties in $M_{A-HT}$ especially for shocks with $\theta_{Bn} \approx 90^\circ$. The large scatter in  $\epsilon/\Delta \Phi$ at same values of $M_{A-HT}$ cannot be fully explained by uncertainties. This indicates that additional physical processes affect  $\epsilon/\Delta \Phi$.

\emph{Conclusions}.-- We analyze suprathermal electron dynamics across 310 quasi-perpendicular collisionless shock crossings with $1.7<M_A<48$ observed by MMS spacecraft using Liouville mapping and the spectral index. On one end,  we find shocks with large non-adiabaticity measure, harder spectra and similar spectral indices in the parallel and perpendicular directions. The spectral index for those shocks is consistent with the prediction of the stochastic shock drift acceleration (SSDA) mechanism \cite{katou2019theory,amano2020observational,amano2022theory}. From this we conclude that the electrons are heated predominantly by a non-adiabatic/stochastic process. Those shocks consist 48\% of the shocks analyzed. As such shocks have harder spectra, it is easier for electrons to be injected into the DSA mechanism \cite{amano2022nonthermal}. On the other end, we find shocks with small non-adiabaticity measure, soft spectra and anisotropic distributions. Those we interpret as shocks with dominant adiabatic behaviour. Statistically the transition between the two types of behavior is controlled by $M_{A-HT}$. For values of $M_{A-HT} \gtrsim30$ we find that stochastic electron dynamics dominate.

MMS data are available at the MMS Science Data Center; see Ref. \footnote{See \url{https://lasp.colorado.edu/mms/sdc/public}.}. Data analysis was performed using the IRFU-MATLAB analysis package \footnote{See \url{https://github.com/irfu/irfu-matlab/}.}.

\emph{Acknowledgement}.-- The authors thank L. Richard for usefull discussions. We thank the MMS team and instrument PIs for data access and support. This work is supported by The Swedish Research Council Grant No. 2018-05514

\bibliography{biblio}

\begin{thebibliography}{46}%
\makeatletter
\providecommand \@ifxundefined [1]{%
 \@ifx{#1\undefined}
}%
\providecommand \@ifnum [1]{%
 \ifnum #1\expandafter \@firstoftwo
 \else \expandafter \@secondoftwo
 \fi
}%
\providecommand \@ifx [1]{%
 \ifx #1\expandafter \@firstoftwo
 \else \expandafter \@secondoftwo
 \fi
}%
\providecommand \natexlab [1]{#1}%
\providecommand \enquote  [1]{``#1''}%
\providecommand \bibnamefont  [1]{#1}%
\providecommand \bibfnamefont [1]{#1}%
\providecommand \citenamefont [1]{#1}%
\providecommand \href@noop [0]{\@secondoftwo}%
\providecommand \href [0]{\begingroup \@sanitize@url \@href}%
\providecommand \@href[1]{\@@startlink{#1}\@@href}%
\providecommand \@@href[1]{\endgroup#1\@@endlink}%
\providecommand \@sanitize@url [0]{\catcode `\\12\catcode `\$12\catcode `\&12\catcode `\#12\catcode `\^12\catcode `\_12\catcode `\%12\relax}%
\providecommand \@@startlink[1]{}%
\providecommand \@@endlink[0]{}%
\providecommand \url  [0]{\begingroup\@sanitize@url \@url }%
\providecommand \@url [1]{\endgroup\@href {#1}{\urlprefix }}%
\providecommand \urlprefix  [0]{URL }%
\providecommand \Eprint [0]{\href }%
\providecommand \doibase [0]{https://doi.org/}%
\providecommand \selectlanguage [0]{\@gobble}%
\providecommand \bibinfo  [0]{\@secondoftwo}%
\providecommand \bibfield  [0]{\@secondoftwo}%
\providecommand \translation [1]{[#1]}%
\providecommand \BibitemOpen [0]{}%
\providecommand \bibitemStop [0]{}%
\providecommand \bibitemNoStop [0]{.\EOS\space}%
\providecommand \EOS [0]{\spacefactor3000\relax}%
\providecommand \BibitemShut  [1]{\csname bibitem#1\endcsname}%
\let\auto@bib@innerbib\@empty
\bibitem [{\citenamefont {Goodrich}\ and\ \citenamefont {Scudder}(1984)}]{goodrich1984adiabatic}%
  \BibitemOpen
  \bibfield  {author} {\bibinfo {author} {\bibfnamefont {C.}~\bibnamefont {Goodrich}}\ and\ \bibinfo {author} {\bibfnamefont {J.}~\bibnamefont {Scudder}},\ }\bibfield  {title} {\bibinfo {title} {The adiabatic energy change of plasma electrons and the frame dependence of the cross-shock potential at collisionless magnetosonic shock waves},\ }\href@noop {} {\bibfield  {journal} {\bibinfo  {journal} {Journal of Geophysical Research: Space Physics}\ }\textbf {\bibinfo {volume} {89}},\ \bibinfo {pages} {6654} (\bibinfo {year} {1984})}\BibitemShut {NoStop}%
\bibitem [{\citenamefont {Schwartz}\ \emph {et~al.}(1988)\citenamefont {Schwartz}, \citenamefont {Thomsen}, \citenamefont {Bame},\ and\ \citenamefont {Stansberry}}]{schwartz1988electron}%
  \BibitemOpen
  \bibfield  {author} {\bibinfo {author} {\bibfnamefont {S.~J.}\ \bibnamefont {Schwartz}}, \bibinfo {author} {\bibfnamefont {M.~F.}\ \bibnamefont {Thomsen}}, \bibinfo {author} {\bibfnamefont {S.}~\bibnamefont {Bame}},\ and\ \bibinfo {author} {\bibfnamefont {J.}~\bibnamefont {Stansberry}},\ }\bibfield  {title} {\bibinfo {title} {Electron heating and the potential jump across fast mode shocks},\ }\href@noop {} {\bibfield  {journal} {\bibinfo  {journal} {Journal of Geophysical Research: Space Physics}\ }\textbf {\bibinfo {volume} {93}},\ \bibinfo {pages} {12923} (\bibinfo {year} {1988})}\BibitemShut {NoStop}%
\bibitem [{\citenamefont {Savoini}\ and\ \citenamefont {Lembege}(1994)}]{savoini1994electron}%
  \BibitemOpen
  \bibfield  {author} {\bibinfo {author} {\bibfnamefont {P.}~\bibnamefont {Savoini}}\ and\ \bibinfo {author} {\bibfnamefont {B.}~\bibnamefont {Lembege}},\ }\bibfield  {title} {\bibinfo {title} {Electron dynamics in two-and one-dimensional oblique supercritical collisionless magnetosonic shocks},\ }\href@noop {} {\bibfield  {journal} {\bibinfo  {journal} {Journal of Geophysical Research: Space Physics}\ }\textbf {\bibinfo {volume} {99}},\ \bibinfo {pages} {6609} (\bibinfo {year} {1994})}\BibitemShut {NoStop}%
\bibitem [{\citenamefont {Scudder}(1995)}]{scudder1995review}%
  \BibitemOpen
  \bibfield  {author} {\bibinfo {author} {\bibfnamefont {J.~D.}\ \bibnamefont {Scudder}},\ }\bibfield  {title} {\bibinfo {title} {A review of the physics of electron heating at collisionless shocks},\ }\href@noop {} {\bibfield  {journal} {\bibinfo  {journal} {Advances in Space Research}\ }\textbf {\bibinfo {volume} {15}},\ \bibinfo {pages} {181} (\bibinfo {year} {1995})}\BibitemShut {NoStop}%
\bibitem [{\citenamefont {Thomsen}\ \emph {et~al.}(1987)\citenamefont {Thomsen}, \citenamefont {Mellott}, \citenamefont {Stansberry}, \citenamefont {Bame}, \citenamefont {Gosling},\ and\ \citenamefont {Russell}}]{thomsen1987strong}%
  \BibitemOpen
  \bibfield  {author} {\bibinfo {author} {\bibfnamefont {M.}~\bibnamefont {Thomsen}}, \bibinfo {author} {\bibfnamefont {M.}~\bibnamefont {Mellott}}, \bibinfo {author} {\bibfnamefont {J.}~\bibnamefont {Stansberry}}, \bibinfo {author} {\bibfnamefont {S.}~\bibnamefont {Bame}}, \bibinfo {author} {\bibfnamefont {J.}~\bibnamefont {Gosling}},\ and\ \bibinfo {author} {\bibfnamefont {C.}~\bibnamefont {Russell}},\ }\bibfield  {title} {\bibinfo {title} {Strong electron heating at the earth's bow shock},\ }\href@noop {} {\bibfield  {journal} {\bibinfo  {journal} {Journal of Geophysical Research: Space Physics}\ }\textbf {\bibinfo {volume} {92}},\ \bibinfo {pages} {10119} (\bibinfo {year} {1987})}\BibitemShut {NoStop}%
\bibitem [{\citenamefont {{Scudder}}\ \emph {et~al.}(1986{\natexlab{a}})\citenamefont {{Scudder}}, \citenamefont {{Mangeney}}, \citenamefont {{Lacombe}}, \citenamefont {{Harvey}},\ and\ \citenamefont {{Aggson}}}]{scudder1986highbeta_b}%
  \BibitemOpen
  \bibfield  {author} {\bibinfo {author} {\bibfnamefont {J.~D.}\ \bibnamefont {{Scudder}}}, \bibinfo {author} {\bibfnamefont {A.}~\bibnamefont {{Mangeney}}}, \bibinfo {author} {\bibfnamefont {C.}~\bibnamefont {{Lacombe}}}, \bibinfo {author} {\bibfnamefont {C.~C.}\ \bibnamefont {{Harvey}}},\ and\ \bibinfo {author} {\bibfnamefont {T.~L.}\ \bibnamefont {{Aggson}}},\ }\bibfield  {title} {\bibinfo {title} {{The resolved layer of a collisionless, high {\ensuremath{\beta}}, supercritical, quasi-perpendicular shock wave. 2. Dissipative fluid electrodynamics}},\ }\href {https://doi.org/10.1029/JA091iA10p11053} {\bibfield  {journal} {\bibinfo  {journal} {\jgr}\ }\textbf {\bibinfo {volume} {91}},\ \bibinfo {pages} {11053} (\bibinfo {year} {1986}{\natexlab{a}})}\BibitemShut {NoStop}%
\bibitem [{\citenamefont {{Scudder}}\ \emph {et~al.}(1986{\natexlab{b}})\citenamefont {{Scudder}}, \citenamefont {{Mangeney}}, \citenamefont {{Lacombe}}, \citenamefont {{Harvey}}, \citenamefont {{Wu}},\ and\ \citenamefont {{Anderson}}}]{scudder1986highbeta_c}%
  \BibitemOpen
  \bibfield  {author} {\bibinfo {author} {\bibfnamefont {J.~D.}\ \bibnamefont {{Scudder}}}, \bibinfo {author} {\bibfnamefont {A.}~\bibnamefont {{Mangeney}}}, \bibinfo {author} {\bibfnamefont {C.}~\bibnamefont {{Lacombe}}}, \bibinfo {author} {\bibfnamefont {C.~C.}\ \bibnamefont {{Harvey}}}, \bibinfo {author} {\bibfnamefont {C.~S.}\ \bibnamefont {{Wu}}},\ and\ \bibinfo {author} {\bibfnamefont {R.~R.}\ \bibnamefont {{Anderson}}},\ }\bibfield  {title} {\bibinfo {title} {{The resolved layer of a collisionless, high {\ensuremath{\beta}}, supercritical, quasi-perpendicular shock wave, 3. Vlasov electrodynamics}},\ }\href {https://doi.org/10.1029/JA091iA10p11075} {\bibfield  {journal} {\bibinfo  {journal} {\jgr}\ }\textbf {\bibinfo {volume} {91}},\ \bibinfo {pages} {11075} (\bibinfo {year} {1986}{\natexlab{b}})}\BibitemShut {NoStop}%
\bibitem [{\citenamefont {Hull}\ \emph {et~al.}(2000)\citenamefont {Hull}, \citenamefont {Scudder}, \citenamefont {Fitzenreiter}, \citenamefont {Ogilvie}, \citenamefont {Newbury},\ and\ \citenamefont {Russell}}]{hull2000electron}%
  \BibitemOpen
  \bibfield  {author} {\bibinfo {author} {\bibfnamefont {A.}~\bibnamefont {Hull}}, \bibinfo {author} {\bibfnamefont {J.}~\bibnamefont {Scudder}}, \bibinfo {author} {\bibfnamefont {R.}~\bibnamefont {Fitzenreiter}}, \bibinfo {author} {\bibfnamefont {K.}~\bibnamefont {Ogilvie}}, \bibinfo {author} {\bibfnamefont {J.}~\bibnamefont {Newbury}},\ and\ \bibinfo {author} {\bibfnamefont {C.}~\bibnamefont {Russell}},\ }\bibfield  {title} {\bibinfo {title} {Electron temperature and de hoffmann-teller potential change across the earth's bow shock: New results from isee 1},\ }\href@noop {} {\bibfield  {journal} {\bibinfo  {journal} {Journal of Geophysical Research: Space Physics}\ }\textbf {\bibinfo {volume} {105}},\ \bibinfo {pages} {20957} (\bibinfo {year} {2000})}\BibitemShut {NoStop}%
\bibitem [{\citenamefont {Feldman}\ \emph {et~al.}(1983)\citenamefont {Feldman}, \citenamefont {Anderson}, \citenamefont {Bame}, \citenamefont {Gary}, \citenamefont {Gosling}, \citenamefont {McComas}, \citenamefont {Thomsen}, \citenamefont {Paschmann},\ and\ \citenamefont {Hoppe}}]{feldman1983electron}%
  \BibitemOpen
  \bibfield  {author} {\bibinfo {author} {\bibfnamefont {W.}~\bibnamefont {Feldman}}, \bibinfo {author} {\bibfnamefont {R.}~\bibnamefont {Anderson}}, \bibinfo {author} {\bibfnamefont {S.}~\bibnamefont {Bame}}, \bibinfo {author} {\bibfnamefont {S.}~\bibnamefont {Gary}}, \bibinfo {author} {\bibfnamefont {J.}~\bibnamefont {Gosling}}, \bibinfo {author} {\bibfnamefont {D.}~\bibnamefont {McComas}}, \bibinfo {author} {\bibfnamefont {M.}~\bibnamefont {Thomsen}}, \bibinfo {author} {\bibfnamefont {G.}~\bibnamefont {Paschmann}},\ and\ \bibinfo {author} {\bibfnamefont {M.}~\bibnamefont {Hoppe}},\ }\bibfield  {title} {\bibinfo {title} {Electron velocity distributions near the earth's bow shock},\ }\href@noop {} {\bibfield  {journal} {\bibinfo  {journal} {Journal of Geophysical Research: Space Physics}\ }\textbf {\bibinfo {volume} {88}},\ \bibinfo {pages} {96} (\bibinfo {year} {1983})}\BibitemShut {NoStop}%
\bibitem [{\citenamefont {Hull}\ \emph {et~al.}(1998)\citenamefont {Hull}, \citenamefont {Scudder}, \citenamefont {Frank}, \citenamefont {Paterson},\ and\ \citenamefont {Kivelson}}]{hull1998electron}%
  \BibitemOpen
  \bibfield  {author} {\bibinfo {author} {\bibfnamefont {A.}~\bibnamefont {Hull}}, \bibinfo {author} {\bibfnamefont {J.}~\bibnamefont {Scudder}}, \bibinfo {author} {\bibfnamefont {L.}~\bibnamefont {Frank}}, \bibinfo {author} {\bibfnamefont {W.}~\bibnamefont {Paterson}},\ and\ \bibinfo {author} {\bibfnamefont {M.}~\bibnamefont {Kivelson}},\ }\bibfield  {title} {\bibinfo {title} {Electron heating and phase space signatures at strong and weak quasi-perpendicular shocks},\ }\href@noop {} {\bibfield  {journal} {\bibinfo  {journal} {Journal of Geophysical Research: Space Physics}\ }\textbf {\bibinfo {volume} {103}},\ \bibinfo {pages} {2041} (\bibinfo {year} {1998})}\BibitemShut {NoStop}%
\bibitem [{\citenamefont {Hull}\ \emph {et~al.}(2001)\citenamefont {Hull}, \citenamefont {Scudder}, \citenamefont {Larson},\ and\ \citenamefont {Lin}}]{hull2001electron}%
  \BibitemOpen
  \bibfield  {author} {\bibinfo {author} {\bibfnamefont {A.}~\bibnamefont {Hull}}, \bibinfo {author} {\bibfnamefont {J.}~\bibnamefont {Scudder}}, \bibinfo {author} {\bibfnamefont {D.}~\bibnamefont {Larson}},\ and\ \bibinfo {author} {\bibfnamefont {R.}~\bibnamefont {Lin}},\ }\bibfield  {title} {\bibinfo {title} {Electron heating and phase space signatures at supercritical, fast mode shocks},\ }\href@noop {} {\bibfield  {journal} {\bibinfo  {journal} {Journal of Geophysical Research: Space Physics}\ }\textbf {\bibinfo {volume} {106}},\ \bibinfo {pages} {15711} (\bibinfo {year} {2001})}\BibitemShut {NoStop}%
\bibitem [{\citenamefont {Lefebvre}\ \emph {et~al.}(2007)\citenamefont {Lefebvre}, \citenamefont {Schwartz}, \citenamefont {Fazakerley},\ and\ \citenamefont {D{\'e}cr{\'e}au}}]{lefebvre2007electron}%
  \BibitemOpen
  \bibfield  {author} {\bibinfo {author} {\bibfnamefont {B.}~\bibnamefont {Lefebvre}}, \bibinfo {author} {\bibfnamefont {S.~J.}\ \bibnamefont {Schwartz}}, \bibinfo {author} {\bibfnamefont {A.~F.}\ \bibnamefont {Fazakerley}},\ and\ \bibinfo {author} {\bibfnamefont {P.}~\bibnamefont {D{\'e}cr{\'e}au}},\ }\bibfield  {title} {\bibinfo {title} {Electron dynamics and cross-shock potential at the quasi-perpendicular earth's bow shock},\ }\href@noop {} {\bibfield  {journal} {\bibinfo  {journal} {Journal of Geophysical Research: Space Physics}\ }\textbf {\bibinfo {volume} {112}} (\bibinfo {year} {2007})}\BibitemShut {NoStop}%
\bibitem [{\citenamefont {Sagdeev}(1966)}]{sagdeev1966cooperative}%
  \BibitemOpen
  \bibfield  {author} {\bibinfo {author} {\bibfnamefont {R.}~\bibnamefont {Sagdeev}},\ }\bibfield  {title} {\bibinfo {title} {Cooperative phenomena and shock waves in collisionless plasmas},\ }\href@noop {} {\bibfield  {journal} {\bibinfo  {journal} {Reviews of plasma physics}\ }\textbf {\bibinfo {volume} {4}},\ \bibinfo {pages} {23} (\bibinfo {year} {1966})}\BibitemShut {NoStop}%
\bibitem [{\citenamefont {Wu}\ \emph {et~al.}(1984)\citenamefont {Wu}, \citenamefont {Winske}, \citenamefont {Zhou}, \citenamefont {Tsai}, \citenamefont {Rodriguez}, \citenamefont {Tanaka}, \citenamefont {Papadopoulos}, \citenamefont {Akimoto}, \citenamefont {Lin}, \citenamefont {Leroy} \emph {et~al.}}]{wu1984microinstabilities}%
  \BibitemOpen
  \bibfield  {author} {\bibinfo {author} {\bibfnamefont {C.}~\bibnamefont {Wu}}, \bibinfo {author} {\bibfnamefont {D.}~\bibnamefont {Winske}}, \bibinfo {author} {\bibfnamefont {Y.}~\bibnamefont {Zhou}}, \bibinfo {author} {\bibfnamefont {S.}~\bibnamefont {Tsai}}, \bibinfo {author} {\bibfnamefont {P.}~\bibnamefont {Rodriguez}}, \bibinfo {author} {\bibfnamefont {M.}~\bibnamefont {Tanaka}}, \bibinfo {author} {\bibfnamefont {K.}~\bibnamefont {Papadopoulos}}, \bibinfo {author} {\bibfnamefont {K.}~\bibnamefont {Akimoto}}, \bibinfo {author} {\bibfnamefont {C.}~\bibnamefont {Lin}}, \bibinfo {author} {\bibfnamefont {M.}~\bibnamefont {Leroy}}, \emph {et~al.},\ }\bibfield  {title} {\bibinfo {title} {Microinstabilities associated with a high mach number, perpendicular bow shock},\ }\href@noop {} {\bibfield  {journal} {\bibinfo  {journal} {Space science reviews}\ }\textbf {\bibinfo {volume} {37}},\ \bibinfo {pages} {63} (\bibinfo {year} {1984})}\BibitemShut {NoStop}%
\bibitem [{\citenamefont {{Papadopoulos}}(1985)}]{papadopoulos1985microinstabilities}%
  \BibitemOpen
  \bibfield  {author} {\bibinfo {author} {\bibfnamefont {K.}~\bibnamefont {{Papadopoulos}}},\ }\bibfield  {title} {\bibinfo {title} {{Microinstabilities and anomalous transport}},\ }\href {https://doi.org/10.1029/GM034p0059} {\bibfield  {journal} {\bibinfo  {journal} {Geophysical Monograph Series}\ }\textbf {\bibinfo {volume} {34}},\ \bibinfo {pages} {59} (\bibinfo {year} {1985})}\BibitemShut {NoStop}%
\bibitem [{\citenamefont {Wilson~III}\ \emph {et~al.}(2014{\natexlab{a}})\citenamefont {Wilson~III}, \citenamefont {Sibeck}, \citenamefont {Breneman}, \citenamefont {Contel}, \citenamefont {Cully}, \citenamefont {Turner}, \citenamefont {Angelopoulos},\ and\ \citenamefont {Malaspina}}]{wilson2014quantified1}%
  \BibitemOpen
  \bibfield  {author} {\bibinfo {author} {\bibfnamefont {L.}~\bibnamefont {Wilson~III}}, \bibinfo {author} {\bibfnamefont {D.}~\bibnamefont {Sibeck}}, \bibinfo {author} {\bibfnamefont {A.}~\bibnamefont {Breneman}}, \bibinfo {author} {\bibfnamefont {O.~L.}\ \bibnamefont {Contel}}, \bibinfo {author} {\bibfnamefont {C.}~\bibnamefont {Cully}}, \bibinfo {author} {\bibfnamefont {D.}~\bibnamefont {Turner}}, \bibinfo {author} {\bibfnamefont {V.}~\bibnamefont {Angelopoulos}},\ and\ \bibinfo {author} {\bibfnamefont {D.}~\bibnamefont {Malaspina}},\ }\bibfield  {title} {\bibinfo {title} {Quantified energy dissipation rates in the terrestrial bow shock: 1. analysis techniques and methodology},\ }\href@noop {} {\bibfield  {journal} {\bibinfo  {journal} {Journal of Geophysical Research: Space Physics}\ }\textbf {\bibinfo {volume} {119}},\ \bibinfo {pages} {6455} (\bibinfo {year} {2014}{\natexlab{a}})}\BibitemShut {NoStop}%
\bibitem [{\citenamefont {Wilson~III}\ \emph {et~al.}(2014{\natexlab{b}})\citenamefont {Wilson~III}, \citenamefont {Sibeck}, \citenamefont {Breneman}, \citenamefont {Contel}, \citenamefont {Cully}, \citenamefont {Turner}, \citenamefont {Angelopoulos},\ and\ \citenamefont {Malaspina}}]{wilson2014quantified2}%
  \BibitemOpen
  \bibfield  {author} {\bibinfo {author} {\bibfnamefont {L.~B.}\ \bibnamefont {Wilson~III}}, \bibinfo {author} {\bibfnamefont {D.~G.}\ \bibnamefont {Sibeck}}, \bibinfo {author} {\bibfnamefont {A.~W.}\ \bibnamefont {Breneman}}, \bibinfo {author} {\bibfnamefont {O.~L.}\ \bibnamefont {Contel}}, \bibinfo {author} {\bibfnamefont {C.}~\bibnamefont {Cully}}, \bibinfo {author} {\bibfnamefont {D.~L.}\ \bibnamefont {Turner}}, \bibinfo {author} {\bibfnamefont {V.}~\bibnamefont {Angelopoulos}},\ and\ \bibinfo {author} {\bibfnamefont {D.~M.}\ \bibnamefont {Malaspina}},\ }\bibfield  {title} {\bibinfo {title} {Quantified energy dissipation rates in the terrestrial bow shock: 2. waves and dissipation},\ }\href@noop {} {\bibfield  {journal} {\bibinfo  {journal} {Journal of Geophysical Research: Space Physics}\ }\textbf {\bibinfo {volume} {119}},\ \bibinfo {pages} {6475} (\bibinfo {year} {2014}{\natexlab{b}})}\BibitemShut {NoStop}%
\bibitem [{\citenamefont {Cargill}\ and\ \citenamefont {Papadopoulos}(1988)}]{cargill1988mechanism}%
  \BibitemOpen
  \bibfield  {author} {\bibinfo {author} {\bibfnamefont {P.}~\bibnamefont {Cargill}}\ and\ \bibinfo {author} {\bibfnamefont {K.}~\bibnamefont {Papadopoulos}},\ }\bibfield  {title} {\bibinfo {title} {A mechanism for strong shock electron heating in supernova remnants},\ }\href@noop {} {\bibfield  {journal} {\bibinfo  {journal} {Astrophysical Journal, Part 2-Letters (ISSN 0004-637X), vol. 329, June 1, 1988, p. L29-L32.}\ }\textbf {\bibinfo {volume} {329}},\ \bibinfo {pages} {L29} (\bibinfo {year} {1988})}\BibitemShut {NoStop}%
\bibitem [{\citenamefont {Hoshino}\ and\ \citenamefont {Shimada}(2002)}]{hoshino2002nonthermal}%
  \BibitemOpen
  \bibfield  {author} {\bibinfo {author} {\bibfnamefont {M.}~\bibnamefont {Hoshino}}\ and\ \bibinfo {author} {\bibfnamefont {N.}~\bibnamefont {Shimada}},\ }\bibfield  {title} {\bibinfo {title} {Nonthermal electrons at high mach number shocks: Electron shock surfing acceleration},\ }\href@noop {} {\bibfield  {journal} {\bibinfo  {journal} {The Astrophysical Journal}\ }\textbf {\bibinfo {volume} {572}},\ \bibinfo {pages} {880} (\bibinfo {year} {2002})}\BibitemShut {NoStop}%
\bibitem [{\citenamefont {Katou}\ and\ \citenamefont {Amano}(2019)}]{katou2019theory}%
  \BibitemOpen
  \bibfield  {author} {\bibinfo {author} {\bibfnamefont {T.}~\bibnamefont {Katou}}\ and\ \bibinfo {author} {\bibfnamefont {T.}~\bibnamefont {Amano}},\ }\bibfield  {title} {\bibinfo {title} {Theory of stochastic shock drift acceleration for electrons in the shock transition region},\ }\href@noop {} {\bibfield  {journal} {\bibinfo  {journal} {The Astrophysical Journal}\ }\textbf {\bibinfo {volume} {874}},\ \bibinfo {pages} {119} (\bibinfo {year} {2019})}\BibitemShut {NoStop}%
\bibitem [{\citenamefont {Amano}\ and\ \citenamefont {Hoshino}(2022)}]{amano2022theory}%
  \BibitemOpen
  \bibfield  {author} {\bibinfo {author} {\bibfnamefont {T.}~\bibnamefont {Amano}}\ and\ \bibinfo {author} {\bibfnamefont {M.}~\bibnamefont {Hoshino}},\ }\bibfield  {title} {\bibinfo {title} {Theory of electron injection at oblique shock of finite thickness},\ }\href@noop {} {\bibfield  {journal} {\bibinfo  {journal} {The Astrophysical Journal}\ }\textbf {\bibinfo {volume} {927}},\ \bibinfo {pages} {132} (\bibinfo {year} {2022})}\BibitemShut {NoStop}%
\bibitem [{\citenamefont {Kamaletdinov}\ \emph {et~al.}(2022)\citenamefont {Kamaletdinov}, \citenamefont {Vasko}, \citenamefont {Artemyev}, \citenamefont {Wang},\ and\ \citenamefont {Mozer}}]{kamaletdinov2022quantifying}%
  \BibitemOpen
  \bibfield  {author} {\bibinfo {author} {\bibfnamefont {S.}~\bibnamefont {Kamaletdinov}}, \bibinfo {author} {\bibfnamefont {I.}~\bibnamefont {Vasko}}, \bibinfo {author} {\bibfnamefont {A.}~\bibnamefont {Artemyev}}, \bibinfo {author} {\bibfnamefont {R.}~\bibnamefont {Wang}},\ and\ \bibinfo {author} {\bibfnamefont {F.}~\bibnamefont {Mozer}},\ }\bibfield  {title} {\bibinfo {title} {Quantifying electron scattering by electrostatic solitary waves in the earth's bow shock},\ }\href@noop {} {\bibfield  {journal} {\bibinfo  {journal} {Physics of Plasmas}\ }\textbf {\bibinfo {volume} {29}} (\bibinfo {year} {2022})}\BibitemShut {NoStop}%
\bibitem [{\citenamefont {Amano}\ \emph {et~al.}(2020)\citenamefont {Amano}, \citenamefont {Katou}, \citenamefont {Kitamura}, \citenamefont {Oka}, \citenamefont {Matsumoto}, \citenamefont {Hoshino}, \citenamefont {Saito}, \citenamefont {Yokota}, \citenamefont {Giles}, \citenamefont {Paterson} \emph {et~al.}}]{amano2020observational}%
  \BibitemOpen
  \bibfield  {author} {\bibinfo {author} {\bibfnamefont {T.}~\bibnamefont {Amano}}, \bibinfo {author} {\bibfnamefont {T.}~\bibnamefont {Katou}}, \bibinfo {author} {\bibfnamefont {N.}~\bibnamefont {Kitamura}}, \bibinfo {author} {\bibfnamefont {M.}~\bibnamefont {Oka}}, \bibinfo {author} {\bibfnamefont {Y.}~\bibnamefont {Matsumoto}}, \bibinfo {author} {\bibfnamefont {M.}~\bibnamefont {Hoshino}}, \bibinfo {author} {\bibfnamefont {Y.}~\bibnamefont {Saito}}, \bibinfo {author} {\bibfnamefont {S.}~\bibnamefont {Yokota}}, \bibinfo {author} {\bibfnamefont {B.}~\bibnamefont {Giles}}, \bibinfo {author} {\bibfnamefont {W.}~\bibnamefont {Paterson}}, \emph {et~al.},\ }\bibfield  {title} {\bibinfo {title} {Observational evidence for stochastic shock drift acceleration of electrons at the earth’s bow shock},\ }\href@noop {} {\bibfield  {journal} {\bibinfo  {journal} {Physical review letters}\ }\textbf {\bibinfo {volume} {124}},\ \bibinfo {pages} {065101} (\bibinfo {year} {2020})}\BibitemShut {NoStop}%
\bibitem [{\citenamefont {Cole}(1976)}]{cole1976effects}%
  \BibitemOpen
  \bibfield  {author} {\bibinfo {author} {\bibfnamefont {K.}~\bibnamefont {Cole}},\ }\bibfield  {title} {\bibinfo {title} {Effects of crossed magnetic and (spatially dependent) electric fields on charged particle motion},\ }\href@noop {} {\bibfield  {journal} {\bibinfo  {journal} {Planetary and Space Science}\ }\textbf {\bibinfo {volume} {24}},\ \bibinfo {pages} {515} (\bibinfo {year} {1976})}\BibitemShut {NoStop}%
\bibitem [{\citenamefont {Balikhin}\ \emph {et~al.}(1998)\citenamefont {Balikhin}, \citenamefont {Krasnosel'skikh}, \citenamefont {Woolliscroft},\ and\ \citenamefont {Gedalin}}]{balikhin1998study}%
  \BibitemOpen
  \bibfield  {author} {\bibinfo {author} {\bibfnamefont {M.}~\bibnamefont {Balikhin}}, \bibinfo {author} {\bibfnamefont {V.}~\bibnamefont {Krasnosel'skikh}}, \bibinfo {author} {\bibfnamefont {L.}~\bibnamefont {Woolliscroft}},\ and\ \bibinfo {author} {\bibfnamefont {M.}~\bibnamefont {Gedalin}},\ }\bibfield  {title} {\bibinfo {title} {A study of the dispersion of the electron distribution in the presence of e and b gradients: Application to electron heating at quasi-perpendicular shocks},\ }\href@noop {} {\bibfield  {journal} {\bibinfo  {journal} {Journal of Geophysical Research: Space Physics}\ }\textbf {\bibinfo {volume} {103}},\ \bibinfo {pages} {2029} (\bibinfo {year} {1998})}\BibitemShut {NoStop}%
\bibitem [{\citenamefont {See}\ \emph {et~al.}(2013)\citenamefont {See}, \citenamefont {Cameron},\ and\ \citenamefont {Schwartz}}]{see2013non}%
  \BibitemOpen
  \bibfield  {author} {\bibinfo {author} {\bibfnamefont {V.}~\bibnamefont {See}}, \bibinfo {author} {\bibfnamefont {R.}~\bibnamefont {Cameron}},\ and\ \bibinfo {author} {\bibfnamefont {S.}~\bibnamefont {Schwartz}},\ }\bibfield  {title} {\bibinfo {title} {Non-adiabatic electron behaviour due to short-scale electric field structures at collisionless shock waves},\ }in\ \href@noop {} {\emph {\bibinfo {booktitle} {Annales Geophysicae}}},\ Vol.~\bibinfo {volume} {31}\ (\bibinfo {organization} {Copernicus Publications G{\"o}ttingen, Germany},\ \bibinfo {year} {2013})\ pp.\ \bibinfo {pages} {639--646}\BibitemShut {NoStop}%
\bibitem [{\citenamefont {Bale}\ \emph {et~al.}(2002)\citenamefont {Bale}, \citenamefont {Hull}, \citenamefont {Larson}, \citenamefont {Lin}, \citenamefont {Muschietti}, \citenamefont {Kellogg}, \citenamefont {Goetz},\ and\ \citenamefont {Monson}}]{bale2002electrostatic}%
  \BibitemOpen
  \bibfield  {author} {\bibinfo {author} {\bibfnamefont {S.}~\bibnamefont {Bale}}, \bibinfo {author} {\bibfnamefont {A.}~\bibnamefont {Hull}}, \bibinfo {author} {\bibfnamefont {D.}~\bibnamefont {Larson}}, \bibinfo {author} {\bibfnamefont {R.}~\bibnamefont {Lin}}, \bibinfo {author} {\bibfnamefont {L.}~\bibnamefont {Muschietti}}, \bibinfo {author} {\bibfnamefont {P.}~\bibnamefont {Kellogg}}, \bibinfo {author} {\bibfnamefont {K.}~\bibnamefont {Goetz}},\ and\ \bibinfo {author} {\bibfnamefont {S.}~\bibnamefont {Monson}},\ }\bibfield  {title} {\bibinfo {title} {Electrostatic turbulence and debye-scale structures associated with electron thermalization at collisionless shocks},\ }\href@noop {} {\bibfield  {journal} {\bibinfo  {journal} {The Astrophysical Journal}\ }\textbf {\bibinfo {volume} {575}},\ \bibinfo {pages} {L25} (\bibinfo {year} {2002})}\BibitemShut {NoStop}%
\bibitem [{\citenamefont {Schwartz}\ \emph {et~al.}(2011)\citenamefont {Schwartz}, \citenamefont {Henley}, \citenamefont {Mitchell},\ and\ \citenamefont {Krasnoselskikh}}]{schwartz2011electron}%
  \BibitemOpen
  \bibfield  {author} {\bibinfo {author} {\bibfnamefont {S.~J.}\ \bibnamefont {Schwartz}}, \bibinfo {author} {\bibfnamefont {E.}~\bibnamefont {Henley}}, \bibinfo {author} {\bibfnamefont {J.}~\bibnamefont {Mitchell}},\ and\ \bibinfo {author} {\bibfnamefont {V.}~\bibnamefont {Krasnoselskikh}},\ }\bibfield  {title} {\bibinfo {title} {Electron temperature gradient scale at collisionless shocks},\ }\href@noop {} {\bibfield  {journal} {\bibinfo  {journal} {Physical review letters}\ }\textbf {\bibinfo {volume} {107}},\ \bibinfo {pages} {215002} (\bibinfo {year} {2011})}\BibitemShut {NoStop}%
\bibitem [{\citenamefont {Wang}\ \emph {et~al.}(2020)\citenamefont {Wang}, \citenamefont {Vasko}, \citenamefont {Mozer}, \citenamefont {Bale}, \citenamefont {Artemyev}, \citenamefont {Bonnell}, \citenamefont {Ergun}, \citenamefont {Giles}, \citenamefont {Lindqvist}, \citenamefont {Russell} \emph {et~al.}}]{wang2020electrostatic}%
  \BibitemOpen
  \bibfield  {author} {\bibinfo {author} {\bibfnamefont {R.}~\bibnamefont {Wang}}, \bibinfo {author} {\bibfnamefont {I.}~\bibnamefont {Vasko}}, \bibinfo {author} {\bibfnamefont {F.}~\bibnamefont {Mozer}}, \bibinfo {author} {\bibfnamefont {S.}~\bibnamefont {Bale}}, \bibinfo {author} {\bibfnamefont {A.}~\bibnamefont {Artemyev}}, \bibinfo {author} {\bibfnamefont {J.}~\bibnamefont {Bonnell}}, \bibinfo {author} {\bibfnamefont {R.}~\bibnamefont {Ergun}}, \bibinfo {author} {\bibfnamefont {B.}~\bibnamefont {Giles}}, \bibinfo {author} {\bibfnamefont {P.-A.}\ \bibnamefont {Lindqvist}}, \bibinfo {author} {\bibfnamefont {C.}~\bibnamefont {Russell}}, \emph {et~al.},\ }\bibfield  {title} {\bibinfo {title} {Electrostatic turbulence and debye-scale structures in collisionless shocks},\ }\href@noop {} {\bibfield  {journal} {\bibinfo  {journal} {The Astrophysical Journal Letters}\ }\textbf {\bibinfo {volume} {889}},\ \bibinfo {pages} {L9} (\bibinfo {year} {2020})}\BibitemShut {NoStop}%
\bibitem [{\citenamefont {Vasko}\ \emph {et~al.}(2022)\citenamefont {Vasko}, \citenamefont {Mozer}, \citenamefont {Bale},\ and\ \citenamefont {Artemyev}}]{vasko2022ion}%
  \BibitemOpen
  \bibfield  {author} {\bibinfo {author} {\bibfnamefont {I.}~\bibnamefont {Vasko}}, \bibinfo {author} {\bibfnamefont {F.}~\bibnamefont {Mozer}}, \bibinfo {author} {\bibfnamefont {S.}~\bibnamefont {Bale}},\ and\ \bibinfo {author} {\bibfnamefont {A.}~\bibnamefont {Artemyev}},\ }\bibfield  {title} {\bibinfo {title} {Ion-acoustic waves in a quasi-perpendicular earth's bow shock},\ }\href@noop {} {\bibfield  {journal} {\bibinfo  {journal} {Geophysical Research Letters}\ }\textbf {\bibinfo {volume} {49}},\ \bibinfo {pages} {e2022GL098640} (\bibinfo {year} {2022})}\BibitemShut {NoStop}%
\bibitem [{\citenamefont {Chen}\ \emph {et~al.}(2018)\citenamefont {Chen}, \citenamefont {Wang}, \citenamefont {Wilson~III}, \citenamefont {Schwartz}, \citenamefont {Bessho}, \citenamefont {Moore}, \citenamefont {Gershman}, \citenamefont {Giles}, \citenamefont {Malaspina}, \citenamefont {Wilder} \emph {et~al.}}]{chen2018electron}%
  \BibitemOpen
  \bibfield  {author} {\bibinfo {author} {\bibfnamefont {L.-J.}\ \bibnamefont {Chen}}, \bibinfo {author} {\bibfnamefont {S.}~\bibnamefont {Wang}}, \bibinfo {author} {\bibfnamefont {L.}~\bibnamefont {Wilson~III}}, \bibinfo {author} {\bibfnamefont {S.}~\bibnamefont {Schwartz}}, \bibinfo {author} {\bibfnamefont {N.}~\bibnamefont {Bessho}}, \bibinfo {author} {\bibfnamefont {T.}~\bibnamefont {Moore}}, \bibinfo {author} {\bibfnamefont {D.}~\bibnamefont {Gershman}}, \bibinfo {author} {\bibfnamefont {B.}~\bibnamefont {Giles}}, \bibinfo {author} {\bibfnamefont {D.}~\bibnamefont {Malaspina}}, \bibinfo {author} {\bibfnamefont {F.}~\bibnamefont {Wilder}}, \emph {et~al.},\ }\bibfield  {title} {\bibinfo {title} {Electron bulk acceleration and thermalization at earth’s quasiperpendicular bow shock},\ }\href@noop {} {\bibfield  {journal} {\bibinfo  {journal} {Physical review letters}\ }\textbf {\bibinfo {volume} {120}},\ \bibinfo {pages} {225101} (\bibinfo {year} {2018})}\BibitemShut {NoStop}%
\bibitem [{\citenamefont {Vasko}\ \emph {et~al.}(2018)\citenamefont {Vasko}, \citenamefont {Mozer}, \citenamefont {Krasnoselskikh}, \citenamefont {Artemyev}, \citenamefont {Agapitov}, \citenamefont {Bale}, \citenamefont {Avanov}, \citenamefont {Ergun}, \citenamefont {Giles}, \citenamefont {Lindqvist} \emph {et~al.}}]{vasko2018solitary}%
  \BibitemOpen
  \bibfield  {author} {\bibinfo {author} {\bibfnamefont {I.}~\bibnamefont {Vasko}}, \bibinfo {author} {\bibfnamefont {F.}~\bibnamefont {Mozer}}, \bibinfo {author} {\bibfnamefont {V.}~\bibnamefont {Krasnoselskikh}}, \bibinfo {author} {\bibfnamefont {A.}~\bibnamefont {Artemyev}}, \bibinfo {author} {\bibfnamefont {O.}~\bibnamefont {Agapitov}}, \bibinfo {author} {\bibfnamefont {S.}~\bibnamefont {Bale}}, \bibinfo {author} {\bibfnamefont {L.}~\bibnamefont {Avanov}}, \bibinfo {author} {\bibfnamefont {R.}~\bibnamefont {Ergun}}, \bibinfo {author} {\bibfnamefont {B.}~\bibnamefont {Giles}}, \bibinfo {author} {\bibfnamefont {P.-A.}\ \bibnamefont {Lindqvist}}, \emph {et~al.},\ }\bibfield  {title} {\bibinfo {title} {Solitary waves across supercritical quasi-perpendicular shocks},\ }\href@noop {} {\bibfield  {journal} {\bibinfo  {journal} {Geophysical Research Letters}\ }\textbf {\bibinfo {volume} {45}},\ \bibinfo {pages} {5809} (\bibinfo {year} {2018})}\BibitemShut {NoStop}%
\bibitem [{\citenamefont {Burch}\ \emph {et~al.}(2016)\citenamefont {Burch}, \citenamefont {Moore}, \citenamefont {Torbert},\ and\ \citenamefont {Giles}}]{MMS_overview}%
  \BibitemOpen
  \bibfield  {author} {\bibinfo {author} {\bibfnamefont {J.}~\bibnamefont {Burch}}, \bibinfo {author} {\bibfnamefont {T.}~\bibnamefont {Moore}}, \bibinfo {author} {\bibfnamefont {R.}~\bibnamefont {Torbert}},\ and\ \bibinfo {author} {\bibfnamefont {B.}~\bibnamefont {Giles}},\ }\bibfield  {title} {\bibinfo {title} {Magnetospheric multiscale overview and science objectives},\ }\href@noop {} {\bibfield  {journal} {\bibinfo  {journal} {Space Science Reviews}\ }\textbf {\bibinfo {volume} {199}},\ \bibinfo {pages} {5} (\bibinfo {year} {2016})}\BibitemShut {NoStop}%
\bibitem [{\citenamefont {Lalti}\ \emph {et~al.}(2022)\citenamefont {Lalti}, \citenamefont {Khotyaintsev}, \citenamefont {Dimmock}, \citenamefont {Johlander}, \citenamefont {Graham},\ and\ \citenamefont {Olshevsky}}]{lalti2022database}%
  \BibitemOpen
  \bibfield  {author} {\bibinfo {author} {\bibfnamefont {A.}~\bibnamefont {Lalti}}, \bibinfo {author} {\bibfnamefont {Y.~V.}\ \bibnamefont {Khotyaintsev}}, \bibinfo {author} {\bibfnamefont {A.}~\bibnamefont {Dimmock}}, \bibinfo {author} {\bibfnamefont {A.}~\bibnamefont {Johlander}}, \bibinfo {author} {\bibfnamefont {D.}~\bibnamefont {Graham}},\ and\ \bibinfo {author} {\bibfnamefont {V.}~\bibnamefont {Olshevsky}},\ }\bibfield  {title} {\bibinfo {title} {A database of mms bow shock crossings compiled using machine learning},\ }\href@noop {} {\bibfield  {journal} {\bibinfo  {journal} {Journal of Geophysical Research: Space Physics}\ }\textbf {\bibinfo {volume} {127}},\ \bibinfo {pages} {e2022JA030454} (\bibinfo {year} {2022})}\BibitemShut {NoStop}%
\bibitem [{\citenamefont {Russell}\ \emph {et~al.}(2016)\citenamefont {Russell}, \citenamefont {Anderson}, \citenamefont {Baumjohann}, \citenamefont {Bromund}, \citenamefont {Dearborn}, \citenamefont {Fischer}, \citenamefont {Le}, \citenamefont {Leinweber}, \citenamefont {Leneman}, \citenamefont {Magnes} \emph {et~al.}}]{russell2016magnetospheric}%
  \BibitemOpen
  \bibfield  {author} {\bibinfo {author} {\bibfnamefont {C.}~\bibnamefont {Russell}}, \bibinfo {author} {\bibfnamefont {B.}~\bibnamefont {Anderson}}, \bibinfo {author} {\bibfnamefont {W.}~\bibnamefont {Baumjohann}}, \bibinfo {author} {\bibfnamefont {K.}~\bibnamefont {Bromund}}, \bibinfo {author} {\bibfnamefont {D.}~\bibnamefont {Dearborn}}, \bibinfo {author} {\bibfnamefont {D.}~\bibnamefont {Fischer}}, \bibinfo {author} {\bibfnamefont {G.}~\bibnamefont {Le}}, \bibinfo {author} {\bibfnamefont {H.}~\bibnamefont {Leinweber}}, \bibinfo {author} {\bibfnamefont {D.}~\bibnamefont {Leneman}}, \bibinfo {author} {\bibfnamefont {W.}~\bibnamefont {Magnes}}, \emph {et~al.},\ }\bibfield  {title} {\bibinfo {title} {The magnetospheric multiscale magnetometers},\ }\href@noop {} {\bibfield  {journal} {\bibinfo  {journal} {Space Science Reviews}\ }\textbf {\bibinfo {volume} {199}},\ \bibinfo {pages} {189} (\bibinfo {year} {2016})}\BibitemShut {NoStop}%
\bibitem [{\citenamefont {Pollock}\ \emph {et~al.}(2016)\citenamefont {Pollock}, \citenamefont {Moore}, \citenamefont {Jacques}, \citenamefont {Burch}, \citenamefont {Gliese}, \citenamefont {Saito}, \citenamefont {Omoto}, \citenamefont {Avanov}, \citenamefont {Barrie}, \citenamefont {Coffey} \emph {et~al.}}]{pollock2016fast}%
  \BibitemOpen
  \bibfield  {author} {\bibinfo {author} {\bibfnamefont {C.}~\bibnamefont {Pollock}}, \bibinfo {author} {\bibfnamefont {T.}~\bibnamefont {Moore}}, \bibinfo {author} {\bibfnamefont {A.}~\bibnamefont {Jacques}}, \bibinfo {author} {\bibfnamefont {J.}~\bibnamefont {Burch}}, \bibinfo {author} {\bibfnamefont {U.}~\bibnamefont {Gliese}}, \bibinfo {author} {\bibfnamefont {Y.}~\bibnamefont {Saito}}, \bibinfo {author} {\bibfnamefont {T.}~\bibnamefont {Omoto}}, \bibinfo {author} {\bibfnamefont {L.}~\bibnamefont {Avanov}}, \bibinfo {author} {\bibfnamefont {A.}~\bibnamefont {Barrie}}, \bibinfo {author} {\bibfnamefont {V.}~\bibnamefont {Coffey}}, \emph {et~al.},\ }\bibfield  {title} {\bibinfo {title} {Fast plasma investigation for magnetospheric multiscale},\ }\href@noop {} {\bibfield  {journal} {\bibinfo  {journal} {Space Science Reviews}\ }\textbf {\bibinfo {volume} {199}},\ \bibinfo {pages} {331} (\bibinfo {year} {2016})}\BibitemShut {NoStop}%
\bibitem [{\citenamefont {{Schwartz}}(1998)}]{schwartz1998}%
  \BibitemOpen
  \bibfield  {author} {\bibinfo {author} {\bibfnamefont {S.~J.}\ \bibnamefont {{Schwartz}}},\ }\bibfield  {title} {\bibinfo {title} {{Shock and Discontinuity Normals, Mach Numbers, and Related Parameters}},\ }\href@noop {} {\bibfield  {journal} {\bibinfo  {journal} {ISSI Scientific Reports Series}\ }\textbf {\bibinfo {volume} {1}},\ \bibinfo {pages} {249} (\bibinfo {year} {1998})}\BibitemShut {NoStop}%
\bibitem [{\citenamefont {Balogh}\ and\ \citenamefont {Treumann}(2013)}]{balogh2013physics}%
  \BibitemOpen
  \bibfield  {author} {\bibinfo {author} {\bibfnamefont {A.}~\bibnamefont {Balogh}}\ and\ \bibinfo {author} {\bibfnamefont {R.~A.}\ \bibnamefont {Treumann}},\ }\href@noop {} {\emph {\bibinfo {title} {Physics of collisionless shocks: space plasma shock waves}}}\ (\bibinfo  {publisher} {Springer Science \& Business Media},\ \bibinfo {year} {2013})\BibitemShut {NoStop}%
\bibitem [{\citenamefont {Schwartz}\ \emph {et~al.}(2021)\citenamefont {Schwartz}, \citenamefont {Ergun}, \citenamefont {Kucharek}, \citenamefont {Wilson~III}, \citenamefont {Chen}, \citenamefont {Goodrich}, \citenamefont {Turner}, \citenamefont {Gingell}, \citenamefont {Madanian}, \citenamefont {Gershman} \emph {et~al.}}]{schwartz2021evaluating}%
  \BibitemOpen
  \bibfield  {author} {\bibinfo {author} {\bibfnamefont {S.~J.}\ \bibnamefont {Schwartz}}, \bibinfo {author} {\bibfnamefont {R.}~\bibnamefont {Ergun}}, \bibinfo {author} {\bibfnamefont {H.}~\bibnamefont {Kucharek}}, \bibinfo {author} {\bibfnamefont {L.}~\bibnamefont {Wilson~III}}, \bibinfo {author} {\bibfnamefont {L.-J.}\ \bibnamefont {Chen}}, \bibinfo {author} {\bibfnamefont {K.}~\bibnamefont {Goodrich}}, \bibinfo {author} {\bibfnamefont {D.}~\bibnamefont {Turner}}, \bibinfo {author} {\bibfnamefont {I.}~\bibnamefont {Gingell}}, \bibinfo {author} {\bibfnamefont {H.}~\bibnamefont {Madanian}}, \bibinfo {author} {\bibfnamefont {D.}~\bibnamefont {Gershman}}, \emph {et~al.},\ }\bibfield  {title} {\bibinfo {title} {Evaluating the dehoffmann-teller cross-shock potential at real collisionless shocks},\ }\href@noop {} {\bibfield  {journal} {\bibinfo  {journal} {Journal of Geophysical Research: Space Physics}\ }\textbf {\bibinfo {volume} {126}},\ \bibinfo {pages} {e2021JA029295} (\bibinfo {year} {2021})}\BibitemShut
  {NoStop}%
\bibitem [{\citenamefont {Johlander}\ \emph {et~al.}(2023)\citenamefont {Johlander}, \citenamefont {Khotyaintsev}, \citenamefont {Dimmock}, \citenamefont {Graham},\ and\ \citenamefont {Lalti}}]{johlander2023electron}%
  \BibitemOpen
  \bibfield  {author} {\bibinfo {author} {\bibfnamefont {A.}~\bibnamefont {Johlander}}, \bibinfo {author} {\bibfnamefont {Y.~V.}\ \bibnamefont {Khotyaintsev}}, \bibinfo {author} {\bibfnamefont {A.~P.}\ \bibnamefont {Dimmock}}, \bibinfo {author} {\bibfnamefont {D.~B.}\ \bibnamefont {Graham}},\ and\ \bibinfo {author} {\bibfnamefont {A.}~\bibnamefont {Lalti}},\ }\bibfield  {title} {\bibinfo {title} {Electron heating scales in collisionless shocks measured by mms},\ }\href@noop {} {\bibfield  {journal} {\bibinfo  {journal} {Geophysical Research Letters}\ }\textbf {\bibinfo {volume} {50}},\ \bibinfo {pages} {e2022GL100400} (\bibinfo {year} {2023})}\BibitemShut {NoStop}%
\bibitem [{sup()}]{supplemental_material}%
  \BibitemOpen
  \href@noop {} {}\bibinfo {note} {See Supplemental Material at [URL will be inserted by publisher] for why the analysis can be done in the SC frame without introducing significant ucnertainty to the results.}\BibitemShut {Stop}%
\bibitem [{\citenamefont {Oka}\ \emph {et~al.}(2006)\citenamefont {Oka}, \citenamefont {Terasawa}, \citenamefont {Seki}, \citenamefont {Fujimoto}, \citenamefont {Kasaba}, \citenamefont {Kojima}, \citenamefont {Shinohara}, \citenamefont {Matsui}, \citenamefont {Matsumoto}, \citenamefont {Saito} \emph {et~al.}}]{oka2006whistler}%
  \BibitemOpen
  \bibfield  {author} {\bibinfo {author} {\bibfnamefont {M.}~\bibnamefont {Oka}}, \bibinfo {author} {\bibfnamefont {T.}~\bibnamefont {Terasawa}}, \bibinfo {author} {\bibfnamefont {Y.}~\bibnamefont {Seki}}, \bibinfo {author} {\bibfnamefont {M.}~\bibnamefont {Fujimoto}}, \bibinfo {author} {\bibfnamefont {Y.}~\bibnamefont {Kasaba}}, \bibinfo {author} {\bibfnamefont {H.}~\bibnamefont {Kojima}}, \bibinfo {author} {\bibfnamefont {I.}~\bibnamefont {Shinohara}}, \bibinfo {author} {\bibfnamefont {H.}~\bibnamefont {Matsui}}, \bibinfo {author} {\bibfnamefont {H.}~\bibnamefont {Matsumoto}}, \bibinfo {author} {\bibfnamefont {Y.}~\bibnamefont {Saito}}, \emph {et~al.},\ }\bibfield  {title} {\bibinfo {title} {Whistler critical mach number and electron acceleration at the bow shock: Geotail observation},\ }\href@noop {} {\bibfield  {journal} {\bibinfo  {journal} {Geophysical Research Letters}\ }\textbf {\bibinfo {volume} {33}} (\bibinfo {year} {2006})}\BibitemShut {NoStop}%
\bibitem [{\citenamefont {Oka}\ \emph {et~al.}(2018)\citenamefont {Oka}, \citenamefont {Birn}, \citenamefont {Battaglia}, \citenamefont {Chaston}, \citenamefont {Hatch}, \citenamefont {Livadiotis}, \citenamefont {Imada}, \citenamefont {Miyoshi}, \citenamefont {Kuhar}, \citenamefont {Effenberger} \emph {et~al.}}]{oka2018electron}%
  \BibitemOpen
  \bibfield  {author} {\bibinfo {author} {\bibfnamefont {M.}~\bibnamefont {Oka}}, \bibinfo {author} {\bibfnamefont {J.}~\bibnamefont {Birn}}, \bibinfo {author} {\bibfnamefont {M.}~\bibnamefont {Battaglia}}, \bibinfo {author} {\bibfnamefont {C.}~\bibnamefont {Chaston}}, \bibinfo {author} {\bibfnamefont {S.}~\bibnamefont {Hatch}}, \bibinfo {author} {\bibfnamefont {G.}~\bibnamefont {Livadiotis}}, \bibinfo {author} {\bibfnamefont {S.}~\bibnamefont {Imada}}, \bibinfo {author} {\bibfnamefont {Y.}~\bibnamefont {Miyoshi}}, \bibinfo {author} {\bibfnamefont {M.}~\bibnamefont {Kuhar}}, \bibinfo {author} {\bibfnamefont {F.}~\bibnamefont {Effenberger}}, \emph {et~al.},\ }\bibfield  {title} {\bibinfo {title} {Electron power-law spectra in solar and space plasmas},\ }\href@noop {} {\bibfield  {journal} {\bibinfo  {journal} {Space Science Reviews}\ }\textbf {\bibinfo {volume} {214}},\ \bibinfo {pages} {82} (\bibinfo {year} {2018})}\BibitemShut {NoStop}%
\bibitem [{\citenamefont {Amano}\ \emph {et~al.}(2022)\citenamefont {Amano}, \citenamefont {Matsumoto}, \citenamefont {Bohdan}, \citenamefont {Kobzar}, \citenamefont {Matsukiyo}, \citenamefont {Oka}, \citenamefont {Niemiec}, \citenamefont {Pohl},\ and\ \citenamefont {Hoshino}}]{amano2022nonthermal}%
  \BibitemOpen
  \bibfield  {author} {\bibinfo {author} {\bibfnamefont {T.}~\bibnamefont {Amano}}, \bibinfo {author} {\bibfnamefont {Y.}~\bibnamefont {Matsumoto}}, \bibinfo {author} {\bibfnamefont {A.}~\bibnamefont {Bohdan}}, \bibinfo {author} {\bibfnamefont {O.}~\bibnamefont {Kobzar}}, \bibinfo {author} {\bibfnamefont {S.}~\bibnamefont {Matsukiyo}}, \bibinfo {author} {\bibfnamefont {M.}~\bibnamefont {Oka}}, \bibinfo {author} {\bibfnamefont {J.}~\bibnamefont {Niemiec}}, \bibinfo {author} {\bibfnamefont {M.}~\bibnamefont {Pohl}},\ and\ \bibinfo {author} {\bibfnamefont {M.}~\bibnamefont {Hoshino}},\ }\bibfield  {title} {\bibinfo {title} {Nonthermal electron acceleration at collisionless quasi-perpendicular shocks},\ }\href@noop {} {\bibfield  {journal} {\bibinfo  {journal} {Reviews of Modern Plasma Physics}\ }\textbf {\bibinfo {volume} {6}},\ \bibinfo {pages} {29} (\bibinfo {year} {2022})}\BibitemShut {NoStop}%
\bibitem [{Note1()}]{Note1}%
  \BibitemOpen
  \bibinfo {note} {See \protect \url {https://lasp.colorado.edu/mms/sdc/public}.}\BibitemShut {Stop}%
\bibitem [{Note2()}]{Note2}%
  \BibitemOpen
  \bibinfo {note} {See \protect \url {https://github.com/irfu/irfu-matlab/}.}\BibitemShut {Stop}%
\end{thebibliography}%

\end{document}


\preprint{APS/123-QED}

\title{Supplemental Material}
\newcommand{\jgr}{Journal of Geophysical Research}

\author{Ahmad Lalti}
\email{ahmad.lalti@irfu.se}
 \affiliation{%
Swedish Institute of Space Physics, Uppsala, Sweden}
\affiliation{Department of Physics and Astronomy, Space and Plasma Physics, Uppsala University, Sweden 
}
\author{Yuri V. Khotyaintsev}
\affiliation{
Swedish Institute of Space Physics, Uppsala, Sweden
}%
\author{Daniel B. Graham}
\affiliation{
Swedish Institute of Space Physics, Uppsala, Sweden
}%

\date{\today}

\maketitle

\section{Equivalency of the analysis of the distributions between the HT and SC frames}

\begin{figure}[h]
    \centering
    \includegraphics[width=8.6cm]{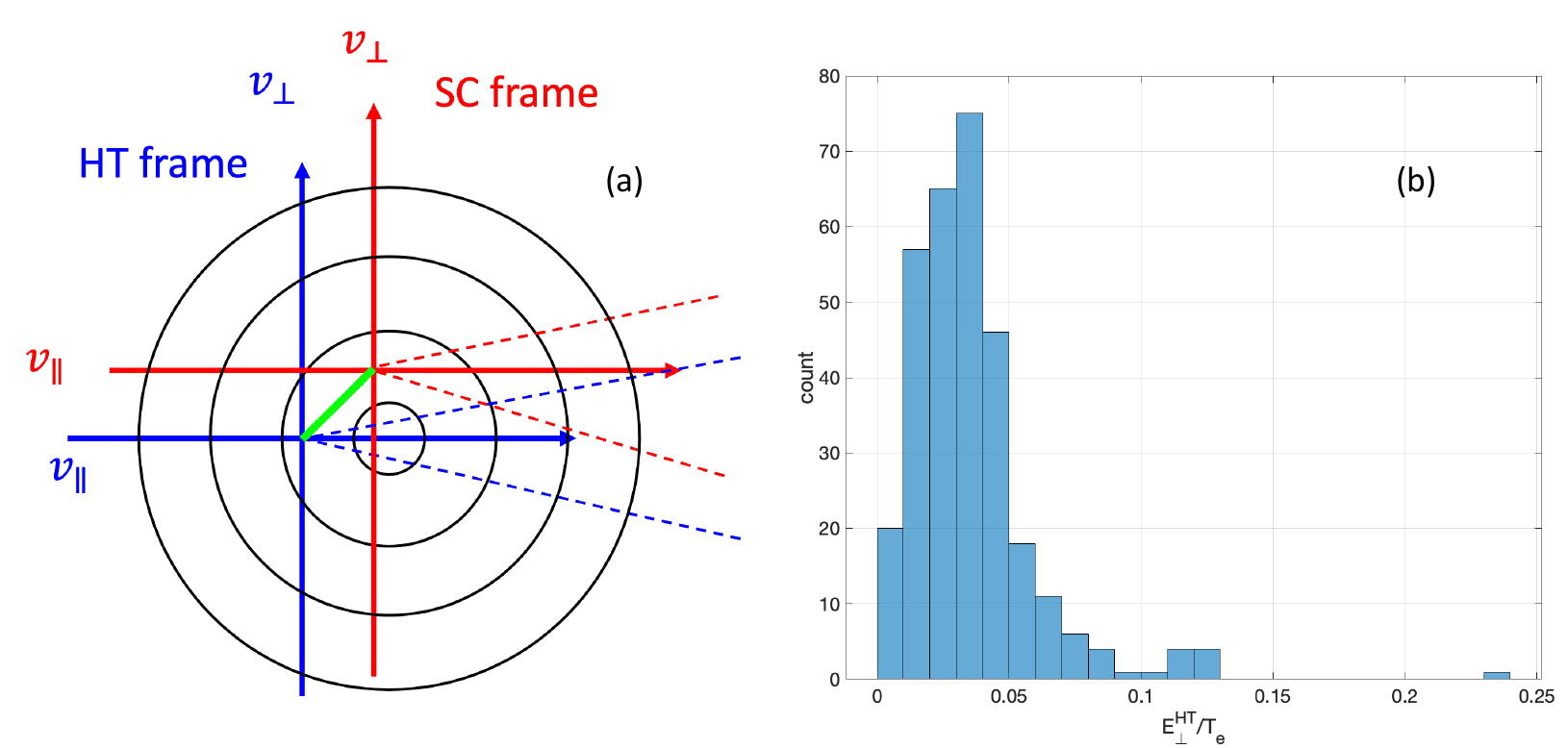}
   \caption{Panel (a) shows schematics of the distribution function in HT (blue) and SC (red) frames. Panel (b) shows histogram of $E_{\perp}^{HT}/T_{e}$.}
   \label{fig:figHT}
\end{figure}

Fig. \ref{fig:figHT} (a) shows a schematics of the same electron distribution function in two different frames of reference the spacecraft (SC) frame in red and the de Hoffmann teller (HT) in blue. The blue and red dashed cones are representation of the parallel cuts in the distribution function in the HT and SC frames respectively. Note that due to the angular resolution of the measured distribution function, electrons with pitch angle $< 15^\circ$ are considered as field aligned.

Ideally one would like to analyze electron distributions in the HT frame (blue dashed cone). But due to the spherical shape of the grid on which the distribution function is discretized along with the logarithmic spacing in energy, such a transformation will introduce large uncertainties.
In the following analysis we show that for the electrons of interest in our analysis (suprathermal electrons with energies at least an order of magnitude larger than the thermal energy) the parallel cuts in the SC frame will be overlapping with those in the HT frame, in other words the red cone will overlap with the blue cone in the schematics of panel (a). 

Let
\begin{equation}
    \mathbf{V}_e^{sc} = V_{e\parallel}^{sc} \hat{b} + V_{e\perp}^{sc}\hat{e}_\perp,
\end{equation}

be the velocity of a suprathermal electron in the spacecraft frame of reference, with $\hat{b}$ and $\hat{e}_\perp$ are the unit vectors parallel and perpendicular to the magnetic field. The pitch angle of that electron in the SC frame is:
\begin{equation}
    \theta^{sc} = \arctan\left(\frac{V_{e\perp}^{sc}}{V_{e\parallel}^{sc}}\right)
    \label{psc}
\end{equation}
Let 
\begin{equation}
    \mathbf{V}_t= \left(V_{b\parallel}^{sc} - \frac{\mathbf{V}_b^{sc}\cdot\hat{n}}{\cos(\theta_{Bn})}\right) \hat{b} + V_{b\perp}^{sc}\hat{e}_\perp,
\end{equation}
be the transformation velocity to go from the SC frame to the HT frame, with $\mathbf{V}_b^{sc} = V_{b\parallel}^{sc} \hat{b} + V_{b\perp}^{sc}\hat{e}_\perp$ being the bulk electron velocity in the SC frame and $\mathbf{n}$ being the shock normal.
So the velocity of the same suprathermal electron in the HT frame can be written as
\begin{equation}
    \mathbf{V}_e^{HT} = \mathbf{V}_e^{sc} - \mathbf{V}_t =\left(V_{e\parallel}^{sc} - V_{b\parallel}^{sc} + \frac{\mathbf{V}_b^{sc}\cdot\hat{n}}{\cos(\theta_{Bn})}\right) \hat{b} + \left(V_{e\perp}^{sc} - V_{b\perp}^{sc}\right)\hat{e}_\perp.
\end{equation}

Subsequently the pitch angle of the suprathermal electron in the HT frame is 
\begin{equation}
    \theta^{HT} = \arctan\left(\frac{\left(V_{e\perp}^{sc} - V_{b\perp}^{sc}\right)}{\left(V_{e\parallel}^{sc} - V_{b\parallel}^{sc} + \frac{\mathbf{V}_b^{sc}\cdot\hat{n}}{\cos(\theta_{Bn})}\right)}\right).
    \label{pHT}
\end{equation}

For electrons with $\theta^{sc} \sim 0^\circ$, from eq. \ref{psc}, $|V_{e\parallel}^{sc}|>>|V_{e\perp}^{sc}|$. If $V_{b\perp}^{sc}<<V_{e\perp}^{sc}$ then $\theta^{HT}$ is also $\sim 0^\circ$, this is because we are considering suprathermal electrons (so $V_{e\parallel}^{sc}>>V_{b\parallel}^{sc}$) and electrons that are moving from upstream to downstream along the field line (so $V_{e\parallel}^{sc}$ and $\frac{\mathbf{V}_b^{sc}\cdot\hat{n}}{\cos(\theta_{Bn})}$ always have the same sign), so the denominator in eq. \ref{pHT} is always larger than the numerator.
If the condition $V_{b\perp}^{sc}<<V_{e\perp}^{sc}$ is satisfied then for the electrons of interest the red cone will overlap with the blue cone in the schemastics of Fig. \ref{fig:figHT} (a).

To test for it we calculate $E_\perp^{HT} = 0.5 m_e \left(V_{b\perp}^{sc}\right)^2$ and normalize it to the electron temperature $T_e$ in energy units for each of our 310 shock, and we plot the histogram of the ratio in Fig. \ref{fig:figHT} (b). We can clearly see that $E_\perp^{HT}/T_e<<1$, and since we are working with suprathermal electrons this means that there is no need to transform to the HT frame and it is sufficient to do the analysis in the SC frame.